# Quantum Entanglement and Measurement Noise: A Novel Approach to Satellite Node Authentication

Pooria Madani[1] and Carolyn McGregor AM[2]

*Ontario Tech University , Oshawa, Ontario, L1G 0C5, Canada, University of Technology Sydney, Ultimo NSW 2007, Australia*

**Abstract**

In this paper, we introduce a novel authentication scheme for satellite nodes based on quantum entanglement and measurement noise profiles. Our approach leverages the unique noise characteristics exhibited by each satellite's quantum optical communication system to create a distinctive "quantum noise fingerprint." This fingerprint is used for node authentication within a satellite constellation, offering a quantum-safe alternative to traditional cryptographic methods. The proposed scheme consists of a training phase, where each satellite engages in a training exercise with its neighbors to compile noise profiles, and an online authentication phase, where these profiles are used for real-time authentication. Our method addresses the inherent challenges of implementing cryptographic-based schemes in space, such as key management and distribution, by exploiting the fundamental properties of quantum mechanics and the unavoidable imperfections in quantum systems. This approach enhances the security and reliability of satellite communication networks, providing a robust solution to the authentication challenges in satellite constellations. We validated and tested several hypotheses for this approach using IBM System One quantum computers.

## I. Introduction

A satellite constellation is a network of satellites working together in a coordinated manner to provide continuous and global coverage for various applications. Unlike a single satellite, a constellation [1] (refer to Fig. 1) comprises multiple satellites strategically positioned in specific orbits to ensure that at least one satellite is always in view of any point on the Earth's surface. This configuration addresses several significant challenges. One primary challenge is providing reliable communication in remote and underserved areas where terrestrial infrastructure is lacking or non-existent [2]. Satellite constellations can bridge this digital divide by delivering high-speed internet and other communication services to these regions [3]. Additionally, they enhance the resilience and redundancy of communication networks. In the event of a satellite failure, other satellites in the constellation can quickly compensate, ensuring minimal service disruption [4]. This redundancy is particularly important for critical applications such as defense, emergency response, and financial transactions.

Today, the space sector is increasingly moving towards building satellite constellations. Advances in satellite technology and miniaturization have made it more feasible and cost-effective to launch and maintain large numbers of satellites. The development of reusable launch vehicles [5] and the reduction in launch costs have also contributed to this trend, enabling more frequent and affordable deployment of satellites. Moreover, the demand for global connectivity and high-speed internet has surged, driven by the proliferation of mobile devices, the Internet of Things (IoT), and the need for seamless communication across different regions.

In addition, the space sector has now expanded to the Space economy [6] given the increasing mission activity by many countries in Cislunar and on the Moon. Cybersecurity of satellites close to the Moon along with spacecraft communication nodes as a proposed architecture proposed in [7] further motivate the need for this work given the challenges of monitoring cybersecurity threat activity in these remote areas of Space. Within that work the need to transmit streaming data to support real-time health and wellness assessment together with equipment health

---

[1] pooria.madani@ontariotechu.ca, Assistant Professor, Faculty of Business and IT, and AIAA Professional Member
[2] carolyn.mcgregor@ontariotechu.ca, Professor, Faculty of Business and IT, Ontario Tech and Professor, Faculty of Engineering and IT, UTS

assessment is an important component in the complex man-instrumentation-equipment-vehicle-environment ecosystem in aerospace missions [8].

Satellite constellations, often comprising dozens to hundreds of satellites and ensuring that each satellite and the data it transmits is authenticated, help prevent unauthorized access, data breaches, and potential malicious attacks that could compromise the entire network. Given the strategic importance of satellite constellations, robust authentication mechanisms are indispensable for maintaining secure and reliable operations. Implementing authentication in satellite-to-satellite communication, however, is a complex and costly task [9]. One of the primary challenges lies in the cryptographic key distribution process. Satellites must be preloaded with cryptographic keys before launch, requiring secure and reliable key management protocols [10]. Once in orbit, synchronizing these keys across the entire constellation is crucial to maintaining secure communications. This synchronization must be achieved despite the vast distances and the dynamic nature of satellite orbits. Furthermore, securely distributing new keys or rotating existing keys poses significant logistical challenges, as any breach in this process could compromise the entire authentication system.

Another layer of complexity is introduced when new satellites join an existing constellation. The addition of new satellites requires updates to the key schedules and storage on all satellites in the network to ensure seamless and secure integration. This process involves generating and distributing new cryptographic keys or updating existing ones, a task that must be performed without disrupting the constellation's operations. The need for continuous key management and synchronization increases the operational complexity and the risk of security vulnerabilities. Ensuring that the entire constellation remains secure during these updates demands robust protocols and meticulous coordination, further escalating the costs and challenges associated with authentication in satellite constellations.

In addition to the stated challenges associated with existing cryptographic schemes, satellites face inherent power and computational resource constraints [11], which further complicate the implementation of cryptographic-based authentication schemes. Unlike terrestrial systems, satellites operate on limited power budgets and have restricted computational capabilities due to their size, weight, and energy constraints [12]. Cryptographic operations, particularly those involving asymmetric algorithms, can be computationally intensive and require significant power [21]. Therefore, designing efficient cryptographic algorithms that provide robust security while minimizing power and computational requirements is essential. This balance is challenging to achieve and often necessitates specialized hardware and software optimizations, which can be costly and time-consuming to develop and implement.

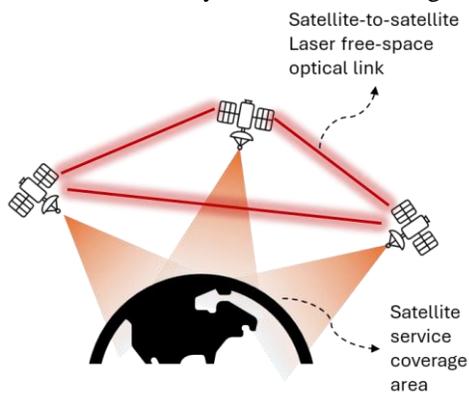

**Fig. 1 A satellite constellation example with satellite-to-satellite communication link.**

Given the substantial challenges associated with cryptographic-based authentication schemes in satellite constellations, it is worthwhile to explore non-cryptographic authentication schemes such as ones that are based on the unique physical characteristics of each satellite. These characteristics, which can include hardware fingerprints [21,22], radiation-induced variations, or manufacturing imperfections, are inherently unique to each satellite and cannot be easily forged or cloned. This uniqueness can provide a robust basis for authentication schemes that are not reliant on traditional cryptographic mechanisms, thereby reducing the computational and logistical burdens associated with key management and distribution in space environments.

In this paper, we propose a novel authentication scheme that leverages the noise profiles exhibited by each satellite in a quantum optical communication setting. Quantum communication systems are inherently sensitive to noise, and the specific noise profile of the quantum optical device onboard a satellite (utilizing optical comm.) can serve as a distinctive identifier. Our novel authentication scheme creates a unique "quantum noise fingerprint" for each satellite in a constellation by analyzing the quantum noise characteristics of the onboard optical communication units. This noise-based authentication method offers a new level of security by utilizing the fundamental and unforgeable properties of quantum noise, providing an innovative solution to the authentication challenges in satellite constellations.

## II. Preliminaries

### A. Quantum Information Representation

A qubit [13], or quantum bit, is the fundamental unit of information in quantum computing and quantum communication. Unlike a classical bit, which can exist in one of two definite states (0 or 1), a qubit can exist simultaneously in a superposition of both states. This unique property arises from the principles of quantum mechanics, particularly superposition, which allows a qubit to be in a state represented as $|\psi\rangle = \alpha|0\rangle + \beta|1\rangle$, where $\alpha$ and $\beta$ are complex numbers that describe the probability amplitudes of the qubit being in states $|0\rangle$ and $|1\rangle$ (use of $|\cdot\rangle$ notation to represent a quantum state is known as Ket notation), respectively. The probabilities $|\alpha|^2$ and $|\beta|^2$ must sum to one, ensuring the qubit is properly normalized. This superposition allows the qubit to hold and process a richer set of information compared to a classical bit, which can only be in one state at a time (either 0 or 1). The ability to exist in superposition is what gives quantum computers and quantum communication systems their potential to perform complex computations and transmit information more efficiently than classical systems.

Entanglement [13,14] is a quantum phenomenon where two or more particles become linked in such a way that the state of one particle instantaneously influences the state of the other, no matter how far apart they are. For example, consider two qubits in an entangled state. A common example is the Bell state [15], represented as $|\psi\rangle = 1/\sqrt{2}(|00\rangle + |11\rangle)$. In this state, the qubits are in a superposition where they are either both in the state $|0\rangle$ or both in the state $|1\rangle$ (i.e., allowed states for a system with two entangled qubits are $|00\rangle$ and $|11\rangle$). If one qubit is measured and found to be in the state $|0\rangle$, the other qubit will immediately be in the state $|0\rangle$ as well, and the same is true if the first qubit is found to be in state $|1\rangle$. This correlation holds regardless of the distance between the qubits, demonstrating the non-local nature of quantum entanglement. Entanglement is crucial for quantum communication protocols such as quantum teleportation and entanglement-based quantum key distribution, and our proposed authentication scheme in this work.

Qubits are implemented using various physical systems, such as the polarization states of photons, the spin states of electrons, or energy levels of atoms. In the context of quantum communication, photons are particularly advantageous due to their ability to travel long distances with minimal interaction with the environment, preserving their quantum states. Quantum communication protocols [16], such as Quantum Key Distribution (QKD) [17], leverage the unique properties of qubits to enable secure communication. In QKD, qubits are used to transmit cryptographic keys. Any attempt to intercept the qubits disturbs their quantum state due to the no-cloning theorem and the principles of quantum measurement, revealing the presence of an eavesdropper and ensuring the security of the communication channel.

The polarization of a photon can be used to represent the binary states 0 and 1 by associating specific polarization directions with each state. In quantum communication and quantum computing, a commonly used convention is to assign the horizontal polarization ($|H\rangle$) to represent the binary state 0 and the vertical polarization ($|V\rangle$) to represent the binary state 1. This means that a photon polarized horizontally can be described by the quantum state $|0\rangle$, and a photon polarized vertically is described by the quantum state $|1\rangle$. Photons can also exist in superpositions of these polarization states, such as diagonal polarization, which can be represented as a combination of horizontal and vertical polarizations, expressed mathematically as $|\psi\rangle = \alpha|H\rangle + \beta|V\rangle$, where α and β are complex probability amplitudes. This use of polarization to encode information allows photons to function as qubits, the fundamental units of information in quantum communication systems, enabling advanced protocols like quantum key distribution and quantum teleportation.

### B. Photon Entanglement

Spontaneous parametric down-conversion (SPDC) [18] is one of the most commonly used techniques for creation of entangled photons. In SPDC, a nonlinear crystal is used to convert a single photon (the pump photon) into a pair of lower-energy photons, called signal and idler photons, which are entangled. This process conserves both energy and momentum, ensuring that the properties of the generated photon pairs are correlated. For example, if the polarization of the pump photon is horizontal, the polarization of the entangled photon pair will be in a superposition of horizontal-horizontal (|HH⟩) and vertical-vertical (|VV⟩) states, resulting in an entangled state like $|\psi\rangle = 1/\sqrt{2}(|HH\rangle + |VV\rangle)$ (important to note that sates |HV⟩ and |VH⟩ will not be detected at the detector in an ideal and noise free setting).

Once entangled, measuring the state of one photon will instantly affect the state of its entangled partner. To measure entangled photons, polarizers or beam splitters along with single-photon detectors are typically used. The measurement setup can be configured to analyze various properties such as polarization, phase, or path. For instance, to measure polarization entanglement, each photon is passed through a polarizer and then detected. The polarizers are set at different angles to test the correlations predicted by quantum mechanics. The results of these measurements can then be used to determine the degree of entanglement.

The correlations between the measurements on the entangled photons are analyzed to reveal quantum mechanical properties that classical particles do not exhibit. For example, measuring one photon in the horizontal polarization state collapses the entangled state, determining the polarization state of the other photon instantaneously. This measurement process verifies entanglement by demonstrating that the measurement outcomes are more strongly correlated than would be possible with classical correlations alone.

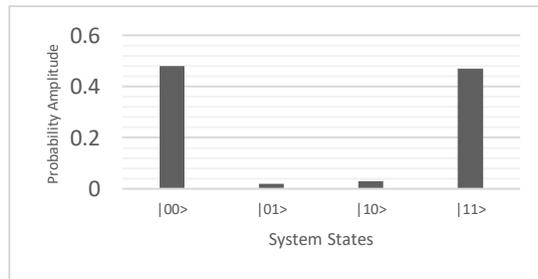

**Fig. 2 This figure illustrates the observed states of a 2-qubit photonic system that are entangled.
Note that states |0⟩ and |H⟩, states |1⟩ and |v⟩ are interchangeably used in this paper.**

Fig. 2 depicts the observed states of a 2-qubit photonic system that are entangled. Ideally, an entangled photon system should predominantly be in the states |00⟩ and |14⟩, and this is reflected in the high probabilities of 0.48 and 0.47 for these states, respectively. However, photon entanglement is not always perfect, and due to various sources of noise and imperfections, such as thermal fluctuations, detector inefficiencies, and environmental disturbances, other states of the system (i.e., violating states), such as |01⟩ and |10⟩, may also be detected at the detector. These states are observed with lower probabilities of 0.02 and 0.03, respectively, indicating the presence of noise and imperfections that affect the purity of the entangled state. Our novel authentication scheme presented in this work uses observed noise distribution over violating states.

**C. Perturbation and Noise**

Entangled states in photonic systems are crucial for various quantum communication and computation applications. However, maintaining the purity and coherence of these entangled states is challenging due to several categories of factors that contribute to noise. These factors can broadly be classified into environmental disturbances, system imperfections, and operational errors.

Environmental disturbances are external factors that can influence the quantum system. Thermal fluctuations are a significant source of noise, as variations in temperature can change the refractive index of optical components and cause phase shifts in the photon states. Electromagnetic interference from surrounding electronic devices or cosmic radiation can also introduce noise, affecting the integrity of the entangled states. These disturbances are often unpredictable and require robust shielding and stabilization techniques to mitigate their impact. As such, this noise evolves over time as a result of ongoing environmental disturbances that can lead to temporal disturbance signatures.

System imperfections arise from the inherent limitations and defects in the components used to create and manipulate entangled photons. For instance, in the process of SPDC, which is commonly used to generate entangled photon pairs, imperfections in the nonlinear crystal or misalignment can lead to the production of additional, uncorrelated photons. Additionally, the detectors used to measure the photon states often have less than 100% efficiency, resulting in photon loss or dark counts—false detection events caused by thermal noise in the detectors. These imperfections reduce the fidelity of the entangled states and introduce errors in the measurement outcomes.

Operational errors occur due to inaccuracies and limitations in the procedures and techniques used to generate, manipulate, and measure the entangled photons. Timing jitter, which refers to the variations in the timing resolution of detectors, can cause uncertainties in identifying coincident photon events, leading to errors in distinguishing between entangled and non-entangled states. Misalignment of optical components, such as beam splitters and polarizers, can also cause deviations from the ideal entangled state. Furthermore, decoherence effects, such as polarization drift caused by birefringence in optical fibers, can degrade the entanglement by mixing the polarization states of the photons.

Our novel authentication scheme capitalizes on these inherent imperfections and noise in entangled photonic systems to develop a unique profile for each quantum transmitter. We hypothesize that these noise profiles, which arise from the enumerated factors are inherently unique and distinguishable across different quantum transmitters. By analyzing the specific noise characteristics, such as variations in phase shifts, detector inefficiencies, and polarization drift, we can create a distinctive "fingerprint" for each quantum transmitter onboard of satellites. This unique noise profile serves as the basis for our authentication scheme, providing a robust and secure method to verify the identity of quantum transmitters in a network, leveraging the unavoidable imperfections in quantum systems to enhance security.

### D. Associated Challenges with Laser and Quantum Communication in Space

Laser optical communication, also known as free-space optical communication (FSO), is an emerging technology that promises to revolutionize satellite-to-satellite communication within satellite constellations. This technology uses laser beams to transmit data between satellites, offering several advantages over traditional radio frequency (RF) communication methods. The adoption of laser optical communication in satellite constellations is driven by its ability to provide higher data bandwidth, reduced latency, and enhanced security, making it an ideal choice for modern space communication networks.

One of the primary benefits of laser optical communication is its ability to operate effectively in the vacuum of space, where there is no atmospheric interference [19]. On Earth, atmospheric conditions such as rain, fog, and turbulence can significantly attenuate and scatter optical signals, reducing their effectiveness for long-distance communication. However, in the near-perfect vacuum of space, these atmospheric disturbances are absent, allowing laser beams to travel vast distances with minimal loss of signal quality. This makes laser communication particularly well-suited for satellite-to-satellite links, where maintaining a clear and stable communication channel is crucial. Moreover, this clear and stable optical transmission environment is ideal for quantum optical communication, which relies on the delicate quantum states of photons (e.g., superposition and entanglement) to encode and transmit information securely.

Moreover, laser optical communication provides improved security and reduced risk of signal interception [19]. Laser beams are highly directional, meaning they can be precisely targeted from one satellite to another. This narrow beam width makes it difficult for unauthorized entities to intercept the signal without being in the direct line of sight between the communicating satellites. In contrast, RF signals tend to spread out over a larger area, making them more susceptible to interception and jamming. The inherent security advantages of laser communication are particularly valuable for military and government applications where secure and reliable communication is paramount.

The existing laser optical communication infrastructure in today's satellite constellations provides a strong foundation for the deployment of future quantum optical communication technologies. Both systems utilize laser technology for data transmission, meaning many of the technical components and engineering principles are already in place. The precision alignment, pointing mechanisms, and high-efficiency detectors developed for classical laser communication can be adapted and optimized for quantum communication purposes [19]. Additionally, the experience and knowledge gained from implementing laser communication systems facilitate the rapid integration and scaling of quantum technologies. This shared infrastructure and expertise significantly reduce the time and cost required to deploy quantum optical communication systems, accelerating their adoption and enhancing the capabilities of satellite constellations.

Developing a fully operational quantum optical communication system for satellite-to-satellite links faces several significant challenges. One of the primary technical hurdles is the precise alignment and stabilization of laser beams between rapidly moving satellites. Quantum communication relies on the transmission of single photons or entangled photon pairs, which requires highly accurate pointing and tracking mechanisms to maintain a stable link. Any misalignment or jitter can lead to a loss of photons, significantly degrading the communication quality and reliability.

Another major challenge is the sensitivity of quantum states to environmental disturbances and noise. Although space provides a near-vacuum environment free from atmospheric interference, other factors such as thermal fluctuations, cosmic radiation, and space weather can still affect the quantum signals. Protecting the delicate quantum states from these disturbances requires sophisticated shielding and error correction techniques, adding complexity to the system design.

Moreover, the generation, detection, and manipulation of quantum states demand advanced and highly sensitive equipment. Single-photon detectors, for instance, must operate with high efficiency and low dark count rates to accurately detect the weak quantum signals. These detectors, along with the necessary quantum sources and modulators, need to be miniaturized and ruggedized to fit within the constraints of satellite platforms while ensuring long-term reliability in the harsh space environment.

## III. Proposed Quantum Authentication Scheme

We hypothesize that the probability distribution formed over the set of $2^n$ states in an $n$-qubit system, after excluding the two allowed entangled states $|0\rangle^n$ and $|1\rangle^n$, is unique to each transmitter. This unique distribution of the remaining states, which we refer to as "error states," can be used for the identification and discrimination of different quantum photonic transmitters within a constellation. Consequently, a recipient, upon receiving a large number of entangled photons of size $n$ can perform measurements and generate the corresponding error distribution. This observed distribution can then be compared to pre-established (i.e., learned) distributions of its peers to verify the identity of the transmitter. In addition, the learned distribution of peers as a set provided by each other node in turn, forms another dimension of a robust way to confirm the authenticity of each node in the constellation, as it not only then has a unique error distribution, but has learned an error distribution of peers in the network.

Let $\vec{X_k} \in \{0,1\}^{2^n}$ denote a vector representing the probability amplitudes of $2^n$ states compiled by making observations/measurements of an $n$-entangled qubits system received $k$ times. Let $y \in \{1,2,\dots,m\}$ denote the set of unique identifiers for $m$ satellite nodes in a constellation. Then, $f^i: X \to y$ is a classifier that the $i^{th}$ satellite node constructs to authenticate and distinguish the $m-1$ other nodes in the constellation by accepting $k$ $n$-entangled photon packets from a transmitting node y.

Our proposed scheme consists of two phases: (a) training and (b) online authentication. In the training phase, each satellite in the constellation will engage in a training exercise with its neighbors. During training, the $i^{th}$ satellite node will query its neighboring satellite $j$ to prepare $k' \gg k$ batches of $n$-entangled photons and optically transmit them to the recipient. The recipient measures the individual received photon batches and uses the results to form $X_{train}^j$. This exercise is repeated with all $m-1$ neighbors. Once the training set is assembled, the node proceeds with training the $f^i$ classification model.

During authentication, satellite $i$ will be probed to engage and authenticate satellite $j$. Node $j$ will prepare $k$ batches of $n$-entangled photons and optically transmit them to node $i$. Upon receiving and measuring these batches to form $X$, the recipient performs authentication by verifying if $f^i(X) = j$.

## IV. Experimental Analysis

We utilized IBM System One quantum computers to validate our proposed authentication scheme and test several hypotheses. While this environment does not replicate the conditions of outer space with optical links, its accessibility and status as a set of real-world quantum computers make it a logical first step. Furthermore, the noise profile of IBM System One during computation closely resembles the noisy conditions that our scheme is designed to operate in. That is, the System One quantum computers each have their own unique system imperfections similarly to computing equipment that would be deployed on a satellite. They are susceptible to operational errors, similarly to space deployments. While the environmental factors are different in this setting, we are still able to demonstrate the unique impact of the environmental and device-specific factors implications in this setting. Thus, this validation provides a solid foundation before investing in the development of infrastructure needed to test the scheme in an outer space environment.

The IBM System One quantum computer with 127 qubits represents a significant milestone in quantum computing technology. It is part of IBM's effort to develop scalable, reliable quantum systems accessible for research and industry applications. With its 127 superconducting qubits, the device can handle more complex quantum computations than its predecessors, enabling advancements in fields like cryptography, optimization, and material science. The architecture of IBM System One integrates state-of-the-art quantum processors with a highly stable environment, minimizing decoherence and noise to improve computation fidelity. This system is also notable for being one of the first quantum computers designed for cloud-based access, allowing users worldwide to run quantum experiments and algorithms remotely. Such capabilities position it as a critical tool for exploring the potential of quantum technology in solving real-world problems.

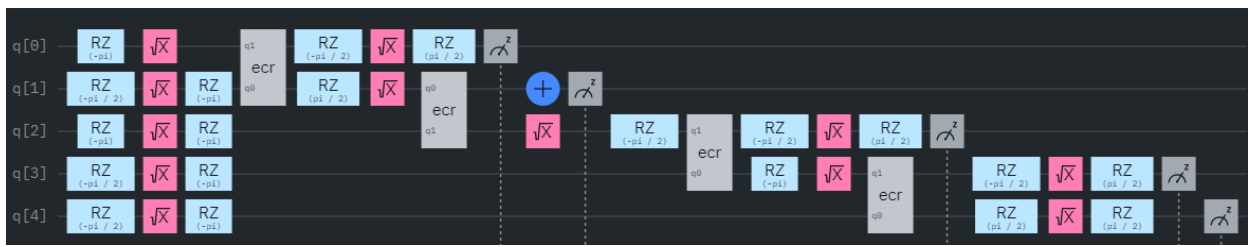

**Fig 3. Qiskit Quantum Circuit implementing our proposed schema in a 5-qubit system. The depicted circuit has been transpiled and optimized by IBM Qiskit's compiler to be executed on IBM System One units.**

In our experiment, we selected four IBM System One quantum computers (names listed in Table 1) equipped with Eagle R3 [20] processors to execute the quantum circuit depicted in Figure 3. This quantum circuit consists of two parts: (a) first, it entangles 5 qubits, and then (b) measures their entangled states. As described in the previous section, such an entangled system can have only two valid solutions upon measurement: the system can collapse into either the state |00000> or |11111>. Due to the stochastic nature of quantum computation, the circuit was executed 10,000 times on each of the selected IBM System One instances to collect the quasi-probability distribution of each computational state, as depicted in Figure 4.

It should be noted that the distributions depicted in Fig. 4 represent quasi-probabilities. Unlike regular (classical) probabilities, which are always non-negative and sum to one, quasi-probabilities can assume negative or even complex values. However, it is possible to transform and normalize these distributions to obtain classical probability distributions, enabling their use in further modeling tasks. Upon visual inspection of the depicted distributions, it becomes evident that incorrect solution states are assigned some probability mass due to quantum computation and environmental noise. Additionally, the shape of these distributions is unique to each quantum computer. However, a technique is required for quantitative computational analysis of the distribution by each satellite.

In other to measure quantitative difference between the collected quasi probability distributions of the four quantum machines with the goal of determining a threshold-value for quantum device identification, we have utilized Kullback-Leilbler (KL) divergence. KL Divergence measures the difference between two probability distributions, typically denoted as $P$ (true distribution) and $Q$ (approximate distribution). It quantifies how much information is lost when $Q$ is used to approximate $P$. Mathematically, it is defined as:

$$D_{KL}(P \parallel Q) = \sum_x P(x) \log \frac{P(x)}{Q(x)}.$$

In the context of our experiment, **P** represents the "learned" distribution of states, encompassing both noisy and correct solutions, to profile a given quantum node. **Q**, on the other hand, represents the distribution of states measured at runtime, which is then compared to **P.** Table 1 highlights clear differences in the measurement results between any two quantum computers executing the quantum circuit depicted in Fig. 3 over 1000 iterations. This preliminary analysis demonstrates promising potential for leveraging noisy solutions in quantum node/device profiling for identification and authentication purposes.

It is important to note that our approach incorporates a classical computer modeling component. Specifically, the size of the discrete state space for measurements in an $n$-qubit system is $2^n$, which grows exponentially with $n$. Capturing the error state probability amplitudes for large values of $n$ (e.g., $n \geq 60$) becomes computationally infeasible due to the sheer size of the state space and associated computational demands. However, even for smaller values of $n$, the system provides a sufficient degree of freedom to effectively accommodate noise profiling for a large number of quantum devices. This is because the smaller state space still captures the critical error characteristics and nuances unique to each quantum device, enabling robust profiling and differentiation across devices.

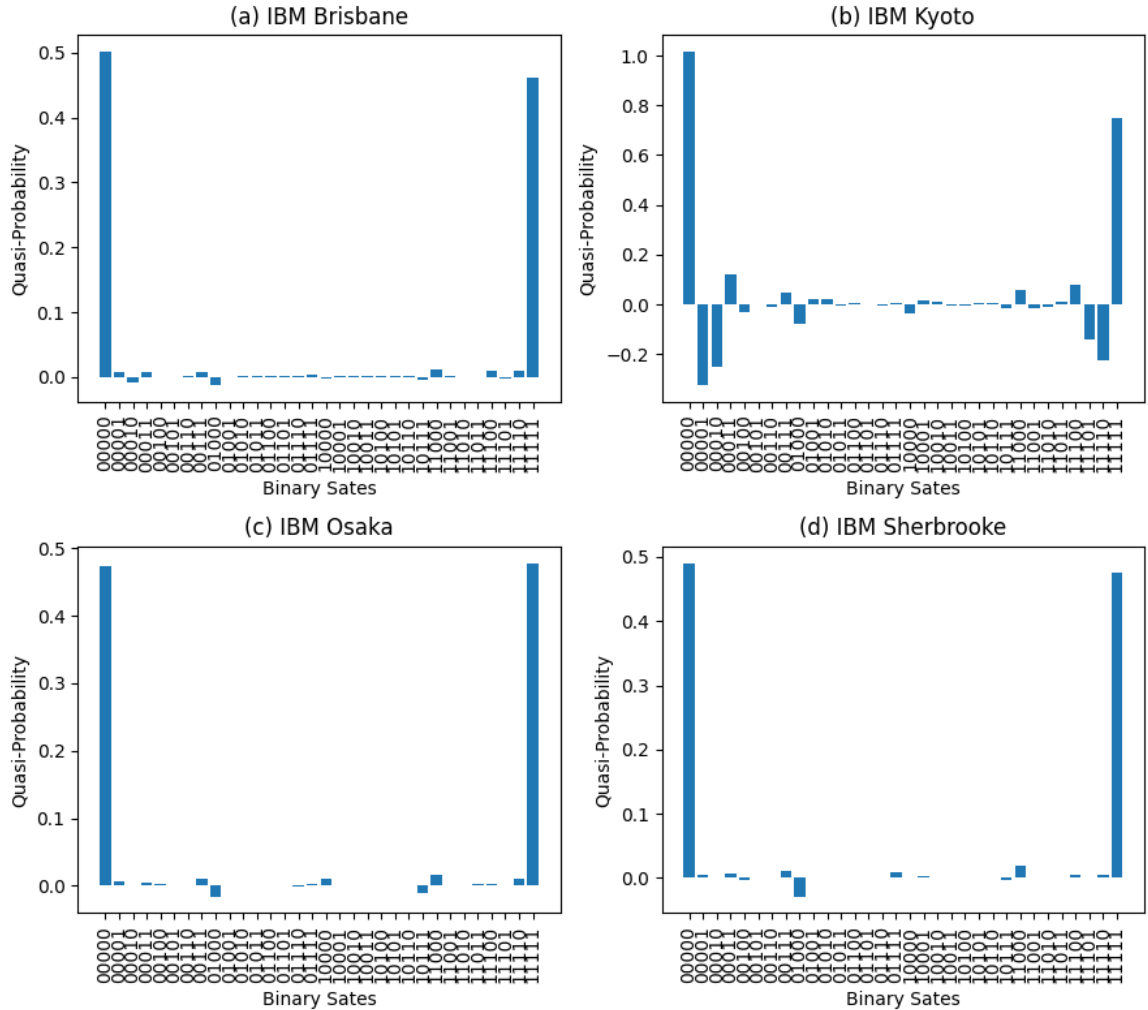

**Fig 4. Execution of the 5-qbit quantum circuit on four different IBM System One 127 Qubits instances. The four graphs illustrate that the correct solution states, |00000> and |11111>, exhibit the highest Quasiprobability. However, various incorrect states, each with different Quasiprobability masses (albeit sometimes marginal), are also present. The distribution of these noisy solutions reveals unique patterns specific to each quantum computer, which can be utilized for identification purposes of the quantum devices.**

It is important to note that environmental disturbances in Space are arguably more subject to change over time due to the extreme environment than the environmental disturbances for the four IBM System One quantum computers we utilised in our experiments. Different radiation profiles for example impacts equipment. Historical review of the degree of change a node is experiencing and the degree to which other nodes are detecting that also adds a further layer for profiling nodes over time.

**Table 1. Kullback-Leibler (KL) divergence computed between pair of quantum computers, based on collected quasi probability distributions from executing the quantum circuit depicted in Fig 3.**

|  | IBM Brisbane | IBM Kyoto | IBM Osaka | IBM Sherbrooke |
|---|---|---|---|---|
| IBM Brisbane | 0.0000 | -0.182081 | 0.067666 | 0.006831 |
| IBM Kyoto | 2.500496 | 0.000000 | 2.512437 | 2.133453 |
| IBM Osaka | 0.282517 | 0.006855 | 0.000000 | 0.085270 |
| IBM Sherbrooke | 0.065112 | -0.305572 | 0.090482 | 0.000000 |

## V. Discussion and Conclusion

In this work, we propose a novel authentication scheme leveraging quantum mechanical properties of light for node authentication between two satellites utilizing quantum computing devices. The proposed scheme is quantum-safe and independent of traditional cryptographic methods. It exploits imperfections during the preparation and measurement of entangled photons. Consequently, any malicious third party attempting to clone the transmitted entangled photons will fail due to the no-cloning theorem, which asserts that the state of a quantum system, such as entangled photons, cannot be cloned. Any cloning attempt would necessitate the regeneration and transmission of new entangled photons by the intermediary, resulting in a substantially different measurement noise profile due to the use of different physical devices compared to the original transmitting (genuine) satellite. We hypothesized that these differences in noise profiles can be robustly detected Within our experiments reported in this paper, we provide initial evidence support of the hypothesis .

One can implement the discriminatory function $f^i$ using various learning-based approaches. The choice of model depends heavily on the available resources within a satellite node, expected concept drift, and the population of nodes in the satellite constellation. Given that $X_k$ represents the probability density function of $2^n$ quantum states, a multinomial distribution may be considered for constructing $f^i$. While fitting a multinomial distribution over a dataset may appear to be an appropriate choice with minimal computational overhead, it is crucial to experimentally verify that this model can reliably discriminate and classify different satellite nodes. Given the abundance of training data, we hypothesize that more complex modelling techniques, such as deep denoising autoencoders (a form of the generative model), may be more suitable and robust for approximating $f^i$.

While it is assumed that the near-vacuum of space provides an ideal environment for implementing the proposed quantum optical communication system, further research is required to assess potential adverse effects from unique outer space disturbances, such as solar winds and cosmic radiation. Moreover, the stability of the noise distribution of each transmitting node remains unclear at the time of this writing. A significant portion of the noise distribution is specific to the transmitting hardware, which can be notably affected by temperature fluctuations common in space, such as those caused by frequent sunrise and sunset cycles. We have attempted to measure the stability of such quantum noise using terrestrial quantum computers provided by IBM System One. The promising preliminary results discussed in this paper provide a strong foundation for further development and testing of our proposed authentication scheme in quantum optical systems. However, in this work, we assumed that the noise profile of quantum computing nodes remains stationary. This assumption needs to be rigorously tested, and it is essential to determine how frequently the noise profile of a quantum computing device changes under varying environmental conditions. Any drift or variation in the noise profile would necessitate the retraining of pre-established noise models, which represents a compelling direction for future research.

In support of the future directions of the new Space economy and real-time monitoring of human health and wellness together with equipment health, we propose future work to test transmissions within the context of the transmission of these forms of data based on the architecture presented in [7].

## References

[1] Kaushal, Hemani, and Georges Kaddoum. "Optical communication in space: Challenges and mitigation techniques." IEEE communications surveys & tutorials 19.1 (2016): 57-96.
[2] Sanad, Ibrahim Shaaban. Reduction of Earth observation system response time using relay satellite constellations. Diss. University of British Columbia, 2020.
[3] Qu, Zhicheng, et al. "LEO satellite constellation for Internet of Things." IEEE access 5 (2017): 18391-18401.

[4] Zeng, Guanming, Yafeng Zhan, and Xiaohan Pan. "Failure-tolerant and low-latency telecommand in mega-constellations: The redundant multi-path routing." IEEE Access 9 (2021): 34975-34985.
[5] Baiocco, P., and Ch Bonnal. "Technology demonstration for reusable launchers." Acta Astronautica 120 (2016): 43-58.
[6] LaRosa, A., McGregor, C., Lord, A., Friend, A., 2024, "Introducing the Interplanetary Chamber of Commerce, 75th International Astronautical Congress, Milan, Italy, 6 pages
[7] McGregor, C., 2021, "A Platform for Real-Time Space Health Analytics as a Service Utilizing Space Data Relays", IEEE Aerospace, Virtual conference, 14 pages
[8] McGregor, C., Orlov, O., Baevsky, R., Chernikova, A., V. Rusanov, "Big Data Analytics for Continuous Assessment of Astronaut Health Risk and Its Application to Human-in-the-Loop (HITL) Related Aerospace Missions", AIAA SciTech Conference, Grapevine, Texas, Jan 2017, 7 pages
[9] von Maurich, Olga, and Alessandro Golkar. "Data authentication, integrity and confidentiality mechanisms for federated satellite systems." Acta Astronautica 149 (2018): 61-76.
[10] Liao, Sheng-Kai, et al. "Satellite-to-ground quantum key distribution." Nature 549.7670 (2017): 43-47.
[11] Saeed, Nasir, et al. "CubeSat communications: Recent advances and future challenges." IEEE Communications Surveys & Tutorials 22.3 (2020): 1839-1862.
[12] Singh, Saurabh, et al. "Advanced lightweight encryption algorithms for IoT devices: survey, challenges and solutions." Journal of Ambient Intelligence and Humanized Computing (2017): 1-18.
[13] Hey, Tony. "Quantum computing: an introduction." Computing & Control Engineering Journal 10.3 (1999): 105-112.
[14] Cleve, Richard, and Harry Buhrman. "Substituting quantum entanglement for communication." Physical Review A 56.2 (1997): 1201.
[15] Kim, Yoon-Ho, Sergei P. Kulik, and Yanhua Shih. "Quantum teleportation of a polarization state with a complete Bell state measurement." Physical Review Letters 86.7 (2001): 1370.
[16] Scully, Marlan O., and M. Suhail Zubairy. Quantum optics. Cambridge university press, 1997.
[17] Scarani, Valerio, et al. "The security of practical quantum key distribution." Reviews of modern physics 81.3 (2009): 1301.
[18] Couteau, Christophe. "Spontaneous parametric down-conversion." Contemporary Physics 59.3 (2018): 291-304.
[19] Kaushal, Hemani, and Georges Kaddoum. "Optical communication in space: Challenges and mitigation techniques." IEEE communications surveys & tutorials 19.1 (2016): 57-96.
[20] Chow, J., Dial, O. and Gambetta, J., 2021. IBM Quantum breaks the 100-qubit processor barrier. IBM Research Blog.
[21] Madani, P. and Vlajic, N., 2021. RSSI-based MAC-layer spoofing detection: deep learning approach. Journal of Cybersecurity and Privacy, 1(3), pp.453-469.
[22] Madani, P., Vlajic, N. and Maljevic, I., 2022. Randomized moving target approach for MAC-layer spoofing detection and prevention in IoT systems. Digital Threats: Research and Practice, 3(4), pp.1-24.